\documentclass[aps,pre,superscriptaddress,amssymb,showpacs,twocolumn]{revtex4-1}
\usepackage{graphicx}
\usepackage{amsmath}
\usepackage{ipa}
\usepackage{amssymb}
\usepackage{makeidx}
\usepackage{comment}
\usepackage{xcolor}

\newcommand{\ket}[1]{\left| #1 \right>} 
\newcommand{\bra}[1]{\left< #1 \right|} 

\def\Tr{\mathrm{Tr}}	 	
\def\diag{\mathrm{diag}}	

\begin{document}

\title{Exact spectral densities of complex noise-plus-structure random matrices}

\author{Jacek Grela} \email{jacekgrela@gmail.com} \affiliation{M. Smoluchowski Institute of Physics and Mark Kac Complex Systems Research Centre, Jagiellonian University,  PL--30348 Krak\'ow, Poland}

\author{Thomas Guhr} \email{thomas.guhr@uni-due.de} \affiliation{Fakult\"at f\"ur Physik,
Universit\"at Duisburg--Essen, Duisburg, Germany}

\begin{abstract}  
  We use supersymmetry to calculate exact spectral densities for a
  class of complex random matrix models having the form $M=S+LXR$,
  where $X$ is a random noise part $X$ and $S,L,R$ are fixed structure
  parts. This is a certain version of the ``external field'' random
  matrix models. We found two--fold integral formulas for arbitrary
  structural matrices. We investigate some special cases in detail and
  carry out numerical simulations. The presence or absence of a
  normality condition on $S$ leads to a qualitatively different
  behavior of the eigenvalue densities.
\end{abstract}

\maketitle


%
%
%
%

\section{Noise-plus-structure random matrices}

In the last 50 years, Random Matrix Theory (RMT) has been established
as an impressively versatile approach \cite{ABP2011:OXFORDHANDBOOK} of
studying complex systems. In particular, applications include large
data structures \cite{QW2013:BIGDATA}, machine learning algorithms
\cite{Ach2004:MACHINELEARNING} and telecommunications
\cite{CD2014:MIMOBOOK} arose recently. It is a common problem in these
and many other areas to infer a signal or information from noisy data.
In this work we study a type of RMT noise-plus-structure model
suitable for this type of inference tasks. More specifically, let $M$
be a matrix of the form:
\begin{align}
\label{meq}
	M = S + L X R,
\end{align}
where $S$ is a fixed matrix and $L,R>0$ are diagonal positive definite
covariance matrices. The matrix $X$ is the source of noise drawn
typically from a multi-dimensional Gaussian ensemble. Equation
\eqref{meq} thus comprises a simplest model combining both randomness
$(X)$ and structure $(S,L,R)$. The matrix $S$ is called a source and
is interpreted as the signal/information matrix of the system in
study. We add a structured noise $LXR$ as every real--world data is
contaminated, and only the resulting matrix $M$ is attainable by
experiment. The matrices $L,R$ encode an anisotropic (or correlated)
source of randomness --- a single element of the source matrix
$S_{ij}$ is perturbed by a noisy term $L_{ii} R_{jj} X_{ij}$,
\textit{i.e.}~with variance $\sigma_{ij}^2 = (L_{ii}R_{jj})^2$.
Absence of any structure means setting $S=0$ and $L=R=1$ which reduces
Eq.~\eqref{meq} to standard RMT models of pure randomness.

There are at least two strategies of studying the model \eqref{meq}
--- we look at either the eigenvalues or the singular values of $M$
(equivalently at the eigenvalues of $M^\dagger M$). The first approach
is limited to square matrices whereas the second route is the main
idea behind the Principal Component Analysis in which, in general,
rectangular data matrices $M$ are investigated. In this work we focus
on the first approach and study the statistics of the eigenvalues. It
is well-known that the symmetries of $M$ constrain the position of its
eigenvalues. Here, however, we drop any symmetry constraints and focus
on the case where eigenvalues spread over the whole complex plane. In
what follows we discuss a couple of instances which can be realized
with the model \eqref{meq} and which are interesting from a practical
as well as from a theoretical perspective.

In finance, one studies the markets to make educated guesses of their
future behaviour, including the search for possibly profitable
correlations. To this end one typically considers $N$ assets in $T$
time slices which may be ordered in a rectangular $N\times T$ matrix
$M$. We set $S=0$ and interpret $L,R$ as noise correlation matrices in
both time and space. Because $M$ is rectangular, the spectral density
of $M^T M$ is studied and thus we arrive at the doubly correlated
Wishart model \cite{WWG2015:DCORRWISHART}. As a second example, in
wireless telecommunication Eq.~\eqref{meq} arises in Multiple
Input Multiple Output (MIMO) systems  as a complex $N_r
\times N_t$ transmission matrix $M$ between $N_t$ transmitters and
$N_r$ receivers \cite{MGC2007:DCORRWISHARTMIMO}.

As a physics application, we consider a Hermitian Hamiltonian $M$
which models an ensemble of charged spinless particles interacting
with a strong external magnetic field \cite{LH1990:TRANSIT}. In this
instance we set $S=e^{-\tau}H_0$, $LR = \sqrt{1-e^{-2\tau}}$ and both
$H_0$ and $X$ are random matrices drawn from the Gaussian Unitary
Ensemble (GUE). The parameter $\tau$ is proportional to the applied
magnetic field. For moderate fields a different Random Matrix Model of
\eqref{meq} applies --- a transition between a Gaussian Orthogonal
Ensemble (GOE) and a GUE happens due to the breaking of time reversal
invariance. In this regime we set $LR = i\alpha$ while the random
matrices $S$ and $X$ are symmetric $S=S^T$ and $X$ antisymmetric
$X=-X^T$, respectively. Even though we drop the positivity
condition of $L,R$ and consider a random matrix $S$, the model
described is still of the form \eqref{meq}. As the parameter $\alpha$ which
is proportional to the field varies between $0 \to 1$, a transition between
GOE and GUE takes place.

Independently, the rich mathematical structure of models of the type
\eqref{meq} has attracted a lot of attention in its own right. These
ensembles are known in the RMT community as ``external source
models''.  So far they were mostly considered for $L=R=1$ and
Hermitian $X$
\cite{BK2004:EXTSOURCE2,BH1998:BHCPAPER2,For2013:BIORTHOGONAL,Guh2006:EXTSOURCE}. These
models also have a natural interpretation in terms of Dyson's Brownian
motion for the stochastic evolution in time $\tau$, when we set
$LR=\sqrt{\tau}$ and view $S$ as the initial matrix
\cite{BGNW2015:HERMDETS,BN2008:CONFTURBPRL}.

Although all of the above examples contain either complex or real
matrices $M$ with a purely real spectrum, there are situations where
symmetry constraints are not present and the spectrum spreads over the
whole complex plane. One of the main tenets of quantum mechanics for
closed systems is the Hermiticity of the Hamiltonian, while dropping
it is an often used effective way to describe open systems,
\textit{i.e.} to account for the environment. As a consequence, complex
energies of the type $E = \varepsilon -i\Gamma$ arise which correspond to
resonant states. Such an energy eigenstate $\ket{\phi_E(t) } =
e^{-iEt} \ket{\phi_E(0)}$ does not only oscillate with a frequency
$\varepsilon$ but also decays with a characteristic time
$1/\Gamma$. Random Matrix Models of this type were used for studying
quantum chaotic scattering in open cavities
\cite{FS1997:RESONANCES2}. In this case, the matrix $S$ is drawn from
the GUE, $LR=-i\pi$, and $X = W^\dagger W$ models a random interaction
between the cavity and its surroundings, where $W$ is drawn from a
complex Girko--Ginibre Ensemble.

As a second application of non--Hermitian matrices, we mention efforts
in constructing mathematical models of neuronal networks
\cite{RA2006:NEURONRMT,LRV1970:RMTNEURO}. Here, $M$ represents the
neuronal adjacency matrix and we begin with setting $S=0,L=R=1$. In
this context however, an additional constraint is needed --- each
matrix row must be either purely negative or purely positive which
reflects Dale's Law of neuronal behaviour. Moreover, a recent paper
\cite{AFM2015:NHCORRSOURCE} argued that also the $S,L$ and $R$
matrices in the model \eqref{meq} might be of significance.

In the sequel, we consider matrices $X$ drawn from the
Girko--Ginibre Ensemble (\textit{i.e.},~a matrix with complex Gaussians
random entries) as well as various types of structural matrices $S,L$ and
$R$. In Sec.~\ref{sec2} we compute an exact formula for the spectral
density of $M$ and arbitrary matrices $S,L$ and $R$. In
Sec.~\ref{sec3} we investigate particular cases: a normal matrix $S$ and
arbitrary matrices $L,R$, a vanishing source $S=0$ and trivial $L=R=1$, and
a rank--one non--normal source $S$ with $L=R=1$. Eventually, we comment on the
spectral formula for a related problem of eigenvalues of
$M^{-1}$. We summarize and conclude in Sec.~\ref{sec4}.

\section{Spectral density of $M$}
\label{sec2}

We now describe the model \eqref{meq} in greater detail. Let $X$ be an
$N\times N$ matrix drawn from a complex Girko--Ginibre Ensemble,
\begin{align}
\label{p}
	P(X) dX = C^{-1} \exp \left ( - n \Tr X^\dagger X \right ) dX,
\end{align}
where $n$ is an (inverse) variance parameter and $C = \left ( \pi/n
\right )^{N^2}$ is the normalization constant. The flat measure over
the matrices $X$ is denoted $dX$. All matrices $S,L$ and $R$ are
$N\times N$, with $L,R$ being positive definite and diagonal. The
source matrix $S$ is in the most general form given by $S=D + T$ where
$D$ is diagonal and $T$ is strictly upper triangular. These reduced
forms are not restrictive because the spectrum of $M$ is unitarily
invariant. In particular, the Schur decomposition of the source matrix
reads $S = U^\dagger (D+T) U$ for a particular unitary matrix
$U$. When $T=0$ the source matrix is called normal, otherwise it is
non-normal.

A basic statistical quantity characterizing the model \eqref{meq} is
the spectral density
\begin{align}
\label{rhodef}
	\rho(z,\bar{z}) = \frac{1}{N} \left <\sum_{i=1}^N \delta^{(2)}(z-m_i) \right >_P,
\end{align}
depending on the complex variable $z$. The $m_i$ are the eigenvalues
of $M$. We use the two--dimensional Dirac delta function due to
complexity of the spectrum, the average is taken over the random
measure \eqref{p}.

Many authors have studied the spectral density \eqref{rhodef} in the
large $N$ limit
\cite{BL2001:NHSPECTR,JN2006:QMETHOD2,BSS2015:NHBROWN}. In particular,
convenient quaternionic/hermitization methods
\cite{JNPZ1997:QMETHOD,FZ1997:HMETHOD} were developed to complete this
task. For $L=R=1$ and a general normal source $S$, spectral density in
the large--$N$ limit was found in Ref.~\cite{Kho1996:NHSPECTR} whereas
the $L,R \neq 1$ generalization was recently studied in
Ref.~\cite{AFM2015:NHCORRSOURCE}. For finite matrix size, a formula
for the spectral density was calculated in
Ref.~\cite{HP1998:NHEXTDENSITY} for $L=R=1$ and a normal source term
$S$ only. In this work we address the cases $L,R \neq 1$ as well as
non--normal $S$.

\subsection{Generating function}

To find the spectral density, we define the averaged ratio of
determinants
\begin{align}
\label{ratiodef}
	\mathcal{R}_{L,R}(Z,V) = \left < \frac{\det ( Z - \mathcal{M} )}{\det ( V - \mathcal{M})} \right >_P
\end{align} 
with the $2N\times 2N$ block matrices
\begin{align}
\mathcal{M} & = \left (
	\begin{matrix}
	0 & M \\
	M^\dagger & 0 \\
	\end{matrix}
	 \right ),  \label{mdef}\\
	 Z & = \left (
	\begin{matrix}
	L^2 w & z \textbf{1}_N \\
	\bar{z} \textbf{1}_N & -R^2\bar{w}\\
	\end{matrix}
	 \right ), V = \left (
	\begin{matrix}
	L^2 u & v \textbf{1}_N \\
	\bar{v} \textbf{1}_N & -R^2\bar{u} \\
	\end{matrix}
	 \right ), \label{zvdef} 
\end{align}
where $\textbf{1}_N$ denotes the $N\times N$ unit matrix. We notice
that the matrices $Z$ and $V$ depend on the complex variables $z$,
$u$, $v$ and $w$. For $u=w=0$ we recover the special case
\begin{align}
	\mathcal{R}_{L,R}(z,v) = \left < \frac{\det [(z-M)(\bar{z}-M^\dagger)]}{\det [(v-M)(\bar{v}-M^\dagger)]}  \right >_{P}.
\end{align}
Although the variables $u,w$ have an interesting interpretation in
terms of the eigenvectors \cite{BGN+2015:NHLONG}, we only use their
regulatory properties -- as long as $u,w \neq 0$, the ratio is finite for all complex $v$.
Importantly, the spectral density is
generated by taking proper derivatives of the averaged ratio, equation
\begin{align}
\label{rho}
	\rho(z,\bar{z}) = -\frac{1}{N\pi} \lim_{\substack{w\to 0}} \frac{\partial}{\partial\bar{z}} \lim_{\substack{V\to Z}} \frac{\partial}{\partial{v}} \mathcal{R}_{L,R}(Z,V)
\end{align}
introduced in Ref.~\cite{FKS1997:WEAKNH} for $L=R=1$. 

As a first step we make the chage of variables $Y = LXR$ implying
$M=S+Y$ as well as $\mathcal{M}=\mathcal{S}+\mathcal{Y}$. The measure
$P(X) dX$ now reads
\begin{align}
	P_{L,R}(Y) dY = C_{L,R}^{-1} \exp \left ( - n \Tr R^{-2} Y^\dagger L^{-2} Y \right ) dY,
\end{align}
where the normalization constant is given as $C_{L,R} = \left ( \pi/n
\right )^{N^2} \det (LR)^2$. We open the ratio of determinants with
the help of complex Grassmann variables $\chi_i$ and complex ordinary
variables $\phi_i$,
\begin{align}
\label{detdetrepr}
	\frac{\det ( Z - \mathcal{M})}{\det ( V - \mathcal{M})} = c \int d[\phi,\chi] e^{ i q^\dagger \diag(V-\mathcal{M},Z-\mathcal{M}) q },
\end{align}
with a proper normalization constant $c$. We introduced the
supervector $q = ( \phi_1 \phi_2 \chi_1 \chi_2 )^T$, and the joint
measure $d[\phi,\chi] = \prod_{i=1}^N d (\phi_1)_i d(\phi_2)_i
d(\chi_1)_i d(\chi_2)_i$. Averaging with the distribution $P_{L,R}$
only affects the exponential terms proportional to $Y$ which are given
by
\begin{align*}
& e^{ -iq^\dagger \diag(\mathcal{Y},\mathcal{Y}) q } = e^{-i\left (\phi_1^\dagger Y \phi_2 + \chi_1^\dagger Y \chi_2 +\phi_2^\dagger Y^\dagger \phi_1 + \chi_2^\dagger Y^\dagger \chi_1 \right )} = \\
 & = e^{- i \Tr ( E_1 Y + E_2 Y^\dagger)},
\end{align*}
where we set $(E_1)_{ij} = (\phi_2)_i (\bar{\phi}_1)_j - (\chi_2)_i
(\bar{\chi}_1)_j$ and $(E_2)_{ij} = (\phi_1)_i (\bar{\phi}_2)_j -
(\chi_1)_i (\bar{\chi}_2)_j$.
The average is easily found to be
\begin{align}
	\int dY P_{L,R}(Y) e^{ - i \Tr ( E_1 Y + E_2 Y^\dagger ) } = e^{ - \frac{1}{n} \Tr E_1 L^2 E_2 R^2 } .
\end{align}
To proceed further, we carry out a Hubbard--Stratonovich transformation 
\begin{align}
	& e^{ - \frac{1}{n} \Tr E_1 L^2 E_2 R^2 } = c_0 \int [d \Sigma] e^{-n F -  q^\dagger Q q},
\label{hs}
\end{align}
which reduces the fourth order supervector terms to second order.
The supermatrix $Q$ appearing in the exponent is given by
\begin{align}
	Q = \left ( \begin{matrix} \mathcal{L} \diag(\sigma\textbf{1}_N,-\bar{\sigma}\textbf{1}_N) & \mathcal{L} \diag(\alpha\textbf{1}_N,\beta\textbf{1}_N) \\ \mathcal{L} \diag(\bar{\alpha}\textbf{1}_N,\bar{\beta}\textbf{1}_N) &  \mathcal{L} \diag(\bar{\rho}\textbf{1}_N,\rho\textbf{1}_N) \end{matrix} \right ) , 
 \end{align} 
with $\mathcal{L} = \diag (L^2,R^2)$. It depends on four new complex
integration variables, two ordinary ones $\sigma$ and $\rho$ as well as
two anticommuting ones $\alpha$ and $\beta$. The corresponding measure
\begin{align}
[d\Sigma] = d^2 \sigma d^2 \rho d^2\alpha d^2\beta
\end{align}
is flat. We use the notation $d^2\alpha=d\alpha d\bar{\alpha}$. The
normalization constant in Eq.~\eqref{hs} is given by $c_0 =
\pi^{-2}$. The function $F = |\sigma|^2 + |\rho|^2 + \bar{\alpha}
\beta + \bar{\beta} \alpha$ in the exponent yield the Gaussians needed
bring the supervector $q$ to second order.

Thus, we can cast the generating function $\mathcal{R}_{L,R}$ into the form
\begin{align}
	\mathcal{R}_{L,R} & = c c_0 \int d [\phi,\chi] \int [d\Sigma] e^{-n F + i q^\dagger A q} ,
\end{align}
where we introduced the supermatrix 
\begin{align}
A = \diag(V-\mathcal{S},Z-\mathcal{S})+ iQ \ .
\end{align}
In the next step we interchange the order of integration $d[\phi,\chi]
\leftrightarrow [d\Sigma]$. This, however, has a subtle flaw: the
resulting integral in the bosonic $\sigma,\rho$ directions is no
longer convergent, an issue addressed previously
\cite{GW1997:SPECTRUMQCDTEMP,JSV1996:UNIVERSALITYNEARZERO}. To
circumvent this problem, we make the change of variables
\begin{align*}
\rho & = \rho_1+i\rho_2, \qquad \sigma=\sigma_1+i\sigma_2, \\
	\rho_1 & = i \frac{w - \bar{w}}{2} + f \cos \phi, \qquad \rho_2 = -\frac{w + \bar{w}}{2} + f \sin \phi, \\
	\sigma_1 & = i\frac{u + \bar{u}}{2} - i g_- \sinh \gamma, \qquad \sigma_2 = \frac{u - \bar{u}}{2} + g_- \cosh \gamma, 
\end{align*}
before swapping the order of integration. Here, we introduced real
commuting variables $f$, $g$, $\gamma$ and $\phi$ as well as a small
imaginary increment, $g_- = g - i \epsilon$ with $\epsilon>0$.  The
range of integration is $f\geq 0,\phi \in (0,2\pi], g\in \mathbb{R},
\gamma\in \mathbb{R}$. The anticommuting variables $\alpha,\beta$
remain unchanged. The integral then becomes
\begin{align}
	\int [d\Sigma] e^{-nF + i q^\dagger A q} = \int [d\Sigma'] (-i g_- f) e^{-nF' + i q^\dagger A' q},
\end{align}
with $[d\Sigma'] = df d\phi dg d\gamma d^2\alpha d^2\beta$ and 
\begin{align}
F' & = g_-^2 + f^2 + |w|^2 - |u|^2 + g_-(u e^\gamma - \bar{u}e^{-\gamma}) \\ \nonumber 
   & \qquad\qquad + if(we^{i\phi}-\bar{w} e^{-i\phi}) + \bar{\alpha} \beta 
                                        + \bar{\beta}\alpha  \nonumber .
\end{align}
We also introduced the transformed supermatrix
\begin{align}
A' = \left ( \begin{matrix} A'_{BB} & A'_{BF} \\ A'_{FB} & A'_{FF} \end{matrix} \right ),
\end{align}
with the $2N\times 2N$ blocks
\begin{align*}
A'_{BB} = & \left ( \begin{matrix} -L^2\sigma_- e^{-s} & v\textbf{1}_N - S \\ \bar{v}\textbf{1}_N - S^\dagger & -R^2\sigma_- e^{s} \end{matrix} \right ) , & A'_{BF} & = \left ( \begin{matrix} i \alpha L^2 & 0 \\ 0 & i \beta R^2 \end{matrix} \right ), \\
A'_{FF} = & \left ( \begin{matrix} i L^2\rho e^{-i\phi} & z\textbf{1}_N - S \\ \bar{z}\textbf{1}_N - S^\dagger & i R^2 \rho e^{i\phi} \end{matrix} \right ), & A'_{FB} & =  \left ( \begin{matrix} i \bar{\alpha} L^2 & 0 \\ 0 & i \bar{\beta} R^2 \end{matrix} \right ).
\end{align*}
After this change of variables, we now may safely interchange the order
of integration and arrive at
\begin{align}
\label{rlr1}
	\mathcal{R}_{L,R} = -i c_0 \int [d\Sigma'] g_- f e^{-nF'} \textrm{sdet}^{-1} A',
\end{align}
where the integral over the supervector yielded the superdeterminant
as an extension of Eq.~\eqref{detdetrepr}
\begin{align}
	c \int d[\phi,\chi] e^{iq^\dagger A' q} = \textrm{sdet}^{-1} A'.
\end{align} 
The superdeterminant is known to satisfy the formula
	 \begin{align*}
	& \textrm{sdet}^{-x} A' = \frac{\det^x A'_{FF}}{\det^x A'_{BB}} \left ( 1 + x \Tr A_0 + \frac{x}{2} \Tr A_0^2 +  \frac{x^2}{2} \left ( \Tr A_0 \right )^2 \right ),
\end{align*}
where $A_0 = A_{BB}'{}^{-1} A_{BF}' A_{FF}'{}^{-1} A_{FB}'$ for any
integer $x$. This result enables us to integrate over the
Grassmann variables $\alpha,\beta$ in Eq.~\eqref{rlr1}. The integral
\begin{align}
I(f,g,\phi,\gamma) = \int d\alpha d\beta e^{-n(\bar{\alpha} \beta + \bar{\beta} \alpha)} \textrm{sdet}^{-1} A' 
\end{align}
can be written in the form
\begin{align}
\label{iintegrated}
	I = -G \left (g_1 +(n-g_2)(n-g_3) + g_4 \right ),
\end{align}
after some algebra and by utilizing the standard normalization of the
Berezin integrals to one. The individual terms are
\begin{align*}
G & = \frac{\det(-f^2\textbf{1}_N - \Gamma_z \Omega_z)}{\det (g_-^2\textbf{1}_N - \Gamma_v \Omega_v)}, \\
	g_2 & = \Tr \left [ \Omega_z \Gamma_v \textbf{P}_v \textbf{Q}_z \right ], \quad g_3 = \Tr \left [ \Omega_v \Gamma_z \mathbf{Q}_z \mathbf{P}_v \right ], \\
	g_1 & = f^2 g_-^2 \Tr \left [ \textbf{P}_v \textbf{Q}_z \right ] \Tr \left [\textbf{P}'_v \textbf{Q}'_z \right ], \\
	g_4 & = f^2 \Tr \left [ \Omega_v \textbf{Q}'_z \Gamma_v \textbf{P}_v \textbf{Q}_z \textbf{P}_v \right ] + g_-^2 \Tr \left [ \Omega_z \textbf{P}'_v \Gamma_z \textbf{Q}_z \textbf{P}_v \textbf{Q}_z \right ],
\end{align*}
where we defined
\begin{align*}
	\Omega_x & = R^{-2} (\bar{x}\textbf{1}_N - S^\dagger), \qquad \Gamma_x = L^{-2}(x\textbf{1}_N-S), \\
	\textbf{P}_v & = (g_-^2\textbf{1}_N - \Omega_v \Gamma_v)^{-1}, \quad \textbf{P}'_v = (g_-^2\textbf{1}_N - \Gamma_v \Omega_v)^{-1}, \\
	\textbf{Q}_z & = (-f^2\textbf{1}_N - \Omega_z \Gamma_z)^{-1}, \quad \textbf{Q}'_z = (-f^2\textbf{1}_N - \Gamma_z\Omega_z)^{-1}.
\end{align*}
At this point we make the remarkable observation that the function $I$
is independent of the variables $\gamma$ and $\phi$ such that
$I(f,g,\phi,\gamma) = I(f,g)$. Hence integrating over the fermionic
variables effectively restores a certain invariance. 

Assembling everything, the generating function \eqref{rlr1} is
given by
\begin{align}
\label{rlr2}
	& \mathcal{R}_{L,R} = -\frac{4i}{\pi} e^{-n|w|^2 + n|u|^2} \int_{-\infty}^\infty dg_- \int_{0}^\infty df J(f,g_-), 
\end{align}
with the integrand  
\begin{align}
J(f,g_-) = g_- f e^{-n(g_-^2 + f^2)} I(f,g_-) I_0(2nf|w|) K_0(2in|u|g_-) ,
\label{int}
\end{align}
depending on the modified Bessel functions $I_0$ and $K_0$ of the
first and second type, respectively. They result
from  the following integrals over the $\gamma,\phi$ variables,
\begin{align*}
 N_\gamma & = \int_{-\infty}^\infty d\gamma e^{-ng_-(u e^\gamma - \bar{u} e^{-\gamma})}, \\
 N_\phi & = \int_0^{2\pi} d\phi e^{-inf(we^{i\phi} - \bar{w} e^{-i\phi})}.
\end{align*}
We set $u = |u| e^{i\theta}, w = |w| e^{i\psi}$ and choose the
argument of $u$ to be $\theta = \pi/2$ to make the $\gamma$ integral
convergent. The angle of $w$ is arbitrary since the $\phi$ integral is
periodic. We therefore set $\psi=0$ and arrive at
\begin{align*}
N_\gamma & = \int_{-\infty}^\infty d\gamma e^{-2ing_- |u| \cosh \gamma} = 2 K_0(2in|u|g_-), \\
N_\phi & = \int_0^{2\pi} d\phi e^{2n f |w| \sin \phi} = 2 \pi I_0(2nf|w|) ,
\end{align*}
which after taking care of the constants yields Eq.~\eqref{rlr2}.

\section{Particular cases}
\label{sec3}

So far, the result \eqref{rlr2} for the generating function is exact
for any matrix dimension $N$ and is valid for any structural matrices
$L,R$ and $S$. Although the integrand \eqref{int} is, in general,
rather complicated, the integral can be worked out explicitly for
certain subclasses of $L,R$ and $S$. We are partcularly interested in
the three cases
\begin{enumerate}
	\item normal source $S$ and variance matrices $L,R$ arbitrary,
	\item vanishing source $S=0$ and trivial $L=R=1$,
	\item non--normal source $S$ of rank one and trivial variance
          matrices $L=R=1$ ,
\end{enumerate}
which we compute and discuss in the sequel.

\subsection{Normal $S$ and arbitrary $L,R$}
\label{secnormal}
In this case all structure matrices $L,R$ and $S$ are diagonal,
\begin{align*}
&	S = \diag (\underbrace{s_1,...,s_1}_{u_1},\underbrace{s_2,...,s_2}_{u_2}, \underbrace{... ,s_x}_{...u_x}), \\
&	L = \diag (\underbrace{l_1,...,l_1}_{v_1},\underbrace{l_2,...,l_2}_{v_2}, \underbrace{... ,l_y}_{...v_y}), \\
&	R = \diag (\underbrace{r_1,...,r_1}_{w_1},\underbrace{r_2,...,r_2}_{w_2}, \underbrace{... ,r_z}_{...w_z}), 
\end{align*}
with three sets of multiplicities $u_i,v_i,w_i$ which should not be
confused with the above employed complex variables $u,v,w$. Here, $x,
y, z$ are the numbers of different entries in the structure matrices
$L,R$ and $S$, respectively, therby defining the sizes of the
sets. The multiplicities in each set add up to $N$. Because the
integrand \eqref{iintegrated} only depends on the products
$(\Omega_x)_{ii} (\Gamma_y)_{ii}$, we introduce a structured source
matrix of the form
\begin{align}
\label{alphadef}
\alpha_{xy} = \Omega_x \Gamma_y = (LR)^{-2} (\bar{x}\textbf{1}_N-S^\dagger)(y\textbf{1}_N-S),
\end{align}
which depends on all three matrices $L,R$ and $S$. It is accompanied
by a merged multiplicity vector $\vec{n}$. We define it by the
following construction: we first form the multiplicity vectors
$\vec{u} = (u_1,...,u_x)$, $\vec{v} = (v_1,...,v_y)$ and $\vec{w} =
(w_1,...,w_z)$ corresponding to the matrices $S, L$ and $R$,
respectively. The vectors $\vec{u}$ is graphically represented by a
column of $N$ points which are ordered in $x$ groups according to the
multiplicities $u_i$. The points within each of these $x$ groups are
given the same (arbitrary) color which is only used to distinguish the
different groups. We refer to the first and last points in each group
as boundary.  The vectors $\vec{v},\vec{w}$ are represented
accordingly.  The multiplicity vector $\vec{n} = (n_1,...,n_k)$ is
then constructed as a vector which has a boundary
whenever \textit{at least} one of the vectors $\vec{u},\vec{v}$ and
$\vec{w}$ has one. We illustrate this by the example in Fig.~1 in
which the vector $\vec{u}$ is represented by $N=11$ points ordered in
$x=3$ groups with multiplicities $u_1=5$, $u_2=2$ and $u_3=4$ with
$5+2+4=11$. As seen, the multiplicities for the other two vectors
differ.  We juxtapose the point sets of all three multiplicity vectors
along with the constructed $\vec{n}$.
\begin{figure}[ht!]
	\centering
	\includegraphics[width=.2\textwidth]{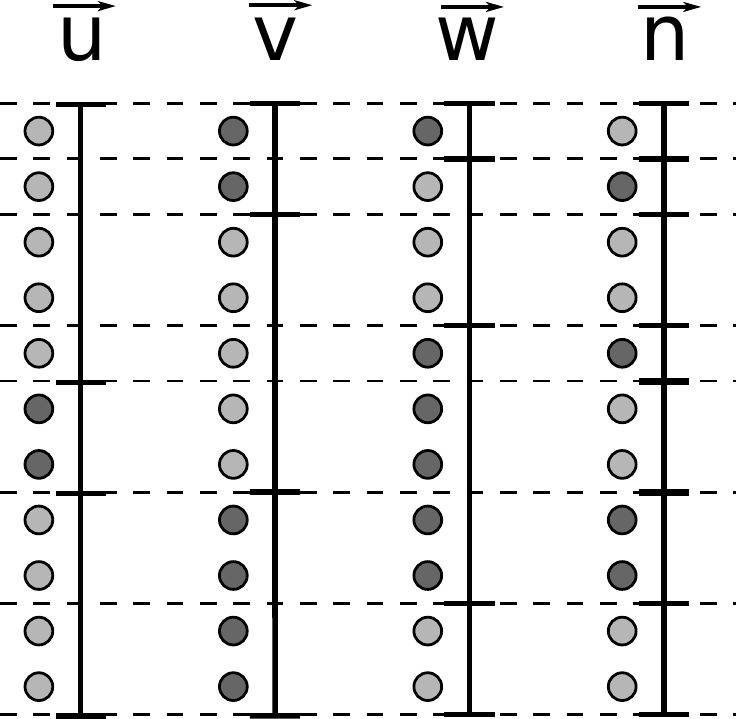}
        \caption{Construction of the multiplicity vector
          $\vec{n}=(1,1,2,1,2,2,2)$ from $\vec{u}=(5,2,4),
          \vec{v}=(2,5,4), \vec{w}=(1,3,5,2)$. The points depict
          groups of sizes determined by the corresponding
          multiplicities. Horizontal lines (both solid and dashed) are
          drawn along the boundaries of the groups of any of the
          vectors $\vec{u},\vec{v}$ and $\vec{w}$, visualizing the
          construction of the merged vector $\vec{n}$.}
\end{figure}
From now on we only use the merged vector $\vec{n}$. We introduce the
dimension $d(\vec{n})$ of the vector $\vec{n}$ as the number of
differing groups, \textit{e.g.} $d(\vec{n})=7$ in the above
example. We also introduce the length $|\vec{n}| = \sum_{i=1}^{d(\vec{n})}n_i$. 
The generating function can then be cast into the form
\begin{align}
\label{normalrlr}
	& \frac{1}{C}\mathcal{R}_{L,R} = i_{\vec{n}} j_{\vec{n}} - \sum_{i=1}^{d(\vec{n})} \frac{n}{n_i} \left (\alpha^{i}_{zv}+\alpha^{i}_{vz} +\frac{N}{n} \right )i_{\vec{n}-\vec{e_i}} j_{\vec{n}+\vec{e_i}} + \nonumber\\
	& + \sum_{i,j=1}^{d(\vec{n})} \frac{n^2 \alpha^i_{zv}}{n_i n_j} \Big [\left ( \alpha^j_{vz} - \alpha^i_{vz} \right ) i_{\vec{n}-\vec{e_i}-\vec{e_j}} j_{\vec{n}+\vec{e_i}+\vec{e_j}} \Big ] + \nonumber\\
	& +\sum_{i,j=1}^{d(\vec{n})} \frac{n}{n_j} \Big [ \alpha^i_{vv} i_{\vec{n}-\vec{e_j}} j_{\vec{n}+\vec{e_i}+\vec{e_j}} + \alpha^i_{zz} i_{\vec{n}-\vec{e_i}-\vec{e_j}} j_{\vec{n}+\vec{e_j}} \Big ], 
\end{align}
where $\alpha^i_{xy}$ is the $i$--th element of the diagonal matrix
\eqref{alphadef}, $C=\prod_{i=1}^{d(\vec{n})} n_i$, and the
$\vec{e_i}$'s are $k$--dimensional unit vectors in the $i$--th
direction. These vectors $\vec{e_i}$ are used to conveniently add or
subtract a single source from the vector $\vec{n}$. The result
\eqref{normalrlr} contains two functions which can be traced back
to the Berezin and the ordinary integrals, We refer to them
as fermionic and as bosonic building blocks. The former is given by 
\begin{align}
\label{idef}
i_{\vec{m}}(z,w) = \frac{e^{-n|w|^2}}{\prod_{i=1}^{{d(\vec{m})}} m_i!} \int_0^\infty d\rho e^{-\rho} I_0(2\sqrt{n\rho} |w|) \prod_{i=1}^{{d(\vec{m})}} \left (\rho + n\alpha^i_{zz} \right )^{m_i},
\end{align}
where we set $i_{\vec{m}} = 0$ if some element of the multiplicity
vector $\vec{m}$ is negative. The bosonic counterpart reads
\begin{align}
\label{jdef0}
& j_{\vec{m}}(v,u) = \frac{2in}{\pi}\prod_{i=1}^{{d(\vec{m})}} \frac{(m_i-1)!}{(-n)^{m_i}} e^{n|u|^2} \times \nonumber \\
& \times \int_{-\infty}^\infty dg g_- e^{-ng_-^2} K_0(2in|u|g_-) \prod_{i=1}^{{d(\vec{m})}}(g_-^2 - \alpha^i_{vv})^{-m_i} .
\end{align}
We notice that the bosonic building block may be expressed as the contour integral
\begin{align}
\label{jdef}
j_{\vec{m}}(v,u) = \frac{\prod_{i=1}^{{d(\vec{m})}} (m_i-1)!}{2\pi i}\oint_{\Gamma_s} dp \sum_{k=0}^\infty \frac{U_{k+1,1}(n|u|^2)p^k}{\prod_{i=1}^{{d(\vec{m})}} \left (p+n \alpha^i_{vv} \right )^{m_i}} ,
\end{align}
where the contour $\Gamma_s$ encircles all sources $-n \alpha^i_{vv}$
counter-clockwise. Here, $U_{a,b}(z) = U(a,b,z)$ is the Tricomi
confluent hypergeometric function. Details of the calculation are
provided in the App.~\ref{appj}.

Before proceeding we cross--check the generating function
\eqref{normalrlr} with similar calculations carry out for the chiral
Gaussian Unitary Ensemble. Choosing the trivial covariance $L=R=1$ and
a vanishing source $S=0$ at the origin $z=v=0$ the generating function
reduces to
\begin{align}
\mathcal{R}_\textrm{chGUE} =\left <\frac{\det(|w|^2 +
    XX^\dagger)}{\det(|u|^2 + XX^\dagger)} \right >_P \ .
\end{align}
We also set $n=N$ and arrive at
\begin{align*}
	\mathcal{R}_\textrm{chGUE} = N \left ( i_N(w) j_N(u) - i_{N-1}(w) j_{N+1}(u) \right ),
\end{align*}
where the index $N$ is a short--hand notation for the one--dimensional
multiplicity vector $\vec{n}=(N)$. We find from the formulas \eqref{idef} and
\eqref{jdef} for the fermionic and bosonic building blocks
\begin{align*}
	i_m(w) = L_m(-N|w|^2), \quad j_m(u) = (x-1)! U_{m,1}(N|u|^2),
\end{align*}
which reproduces the results of Ref.~\cite{FS2002:CHIRALGUECORR}.
However, in the present study we are interested in the complementary
limit, \textit{i.e.}, we set $u,w\to 0$ and look at $z,v \neq 0$.

We now wish to calculate the spectral density. We recall the formula
\eqref{rho} where the parameters $u$ and $w$ serve as regulators. It is
desirable to set them to zero before computing the derivatives. Even
though this does not pose a problem for the fermionic block
\eqref{idef}, it turns out to produce infinities in the bosonic block
\eqref{jdef}. To control these emerging singularities, we use the
identity 
\begin{align*}
 k! U_{k+1,1}(n|u|^2) = & e^{n|u|^2} \Gamma(0,n|u|^2) L_{k}(-n|u|^2)  + \\
 & + \tilde{L}_{k}(-n|u|^2),
\label{gamsing}
\end{align*}
for the confluent hypergeometric function. Here, $L_k$ are the
Laguerre polynomials whereas $\tilde{L}_k$ are defined by the same
recurrence relations but with different initial conditions
$\tilde{L}_0(x) = 0, \tilde{L}_1(x) = -1$. The singular behavior for
$U$ as $u \to 0$ is due to the incomplete Gamma function
$\Gamma(0,n|u|^2)$ in the first term. We therefore split the
bosonic block into a singular and a regular parts,
\begin{align}
  j_{\vec{m}}(v,u) = j^\textrm{(sing)}_{\vec{m}}(v,u) +
  j^\textrm{(reg)}_{\vec{m}}(v,u).
\end{align}
To control the singularity, we set the singular part
$j^\textrm{(sing)}$ to zero and take the limit $u \to 0$ in the
regular part $j^\textrm{(reg)}$. We formalize this procedure by
introducing the regularized generating function
\begin{align}
\label{rlrreg}
	\tilde{\mathcal{R}}_{L,R} = \mathcal{R}_{L,R} \left [ i_{\vec{m}}(z,w) \to \tilde{i}_{\vec{m}}(z), j_{\vec{m}}(v,u) \to \tilde{j}_{\vec{m}}(v) \right ],
\end{align}
with new building blocks $\tilde{i}_{\vec{m}}(z)=i_{\vec{m}}(z,w=0)$
and $\tilde{j}_{\vec{m}}(v) = j^{(reg)}_{\vec{m}}(v,u=0)$ already in
the $w,u\to 0$ limit.  We stress that this procedure is not an
approximation --- although we have $\tilde{\mathcal{R}}_{L,R} \neq
\mathcal{R}_{L,R}$, the spectral densities obtained by Eq.~\eqref{rho}
agree exactly $\tilde{\rho} = \rho$. We checked this numerically.
This property is intuitively justified since we subtract the otherwise
infinite part proportional to $j^\textrm{(sing)}$. The regularized
building blocks are given by
\begin{align}
\label{ijreg}
	\tilde{i}_{\vec{m}} & = \frac{1}{\prod_{i=1}^{d(\vec{m})} m_i!} \int_0^\infty d\rho e^{-\rho} \prod_{i=1}^{d(\vec{m})} \Big (\rho+n\alpha^i_{zz} \Big )^{m_i}, \nonumber \\
	\tilde{j}_{\vec{m}} & = - \frac{\prod_{i=1}^{d(\vec{m})} (m_i-1)!}{2\pi i} \oint_{\Gamma_s} dp \frac{e^p(\gamma+\Gamma(0,p)+\ln p)}{\prod_{i=1}^{d(\vec{m})} (p + n \alpha^i_{vv})^{m_i}},
\end{align}
where we used the identity 
\begin{align}
\sum_{m=0}^\infty \frac{1}{m!} \tilde{L}_{m}(0) p^m = -
e^p (\gamma + \Gamma(0,p)+\ln p)
\end{align}
for the modified Laguerre polynomials with $\gamma$ denoting the Euler
constant. This identity follows from the fact that $\tilde{L}_m(0) =
-\sum_{k=1}^m \frac{1}{k}$ are the (negative) harmonic numbers.

The final formula for the spectral density in the case of a normal source
$S$ and nontrivial $L,R$ then reads
\begin{align}
\label{normalrho}
\tilde{\rho} = - \frac{1}{N\pi} \frac{\partial}{\partial\bar{z}} \lim_{\substack{V\to Z}} \frac{\partial}{\partial{v}} \tilde{\mathcal{R}}_{L,R}(z,v),
\end{align}
together with the definitions \eqref{normalrlr}, \eqref{rlrreg} and
\eqref{ijreg}.  We demonstrate the utility of our analytical result in
Fig.~2 by comparing it with numerical simulations.
\begin{figure}[ht!]
	\centering
	\includegraphics[width=.42\textwidth]{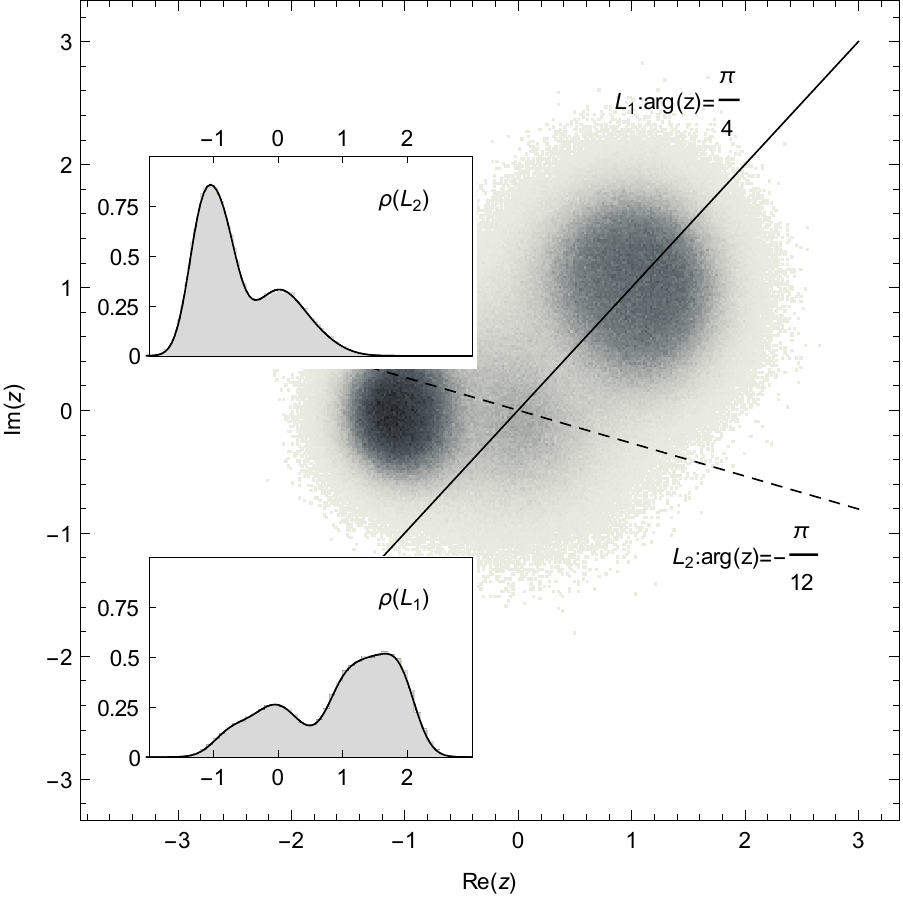}
        \caption{Spectral density according to Eq.~\eqref{normalrho}
          as insets along two lines $L_1$ and $L_2$ in the complex
          plane, together with numerical simulations. The structural
          matrices are $S=\diag(-1,0,1+i)$, $L=\diag(3/4,1)$ and
          $R=\diag(1,5/4,1)$ with multiplicity vectors of
          $\vec{u}=(2,1,3)$, $\vec{v}=(2,4)$ and $\vec{w}=(2,1,3)$.}
\end{figure}
Adding (structured) noise $LXR$ produces an overall eigenvalues
spreading with anisotropic features reflecting the $L,R$ covariance
matrices. The density is concentrated around the initial eigenvalues
of $S$ and varies smoothly as we change the noise level $n$, \textit{i.e.}
the inverse variance of the ensemble~\eqref{p}.

\subsection{Vanishing source $S=0$ and $L=R=1$}
\label{ginsection}

We now consider the case $S=0$ and $L=R=1$ in which a simple spectral
density formula is known from the work of Ginibre
\cite{Gin1965:ENSEMBLE}. The multiplicity vector is one--dimensional
$\vec{n} = (N)$ and the source matrix has the simple form $\alpha_{xy}
= \bar{x} y \textbf{1}_N$. The regularized generating
function~\eqref{rlrreg} reads
\begin{align}
\tilde{\mathcal{R}}_{G} = & N (\tilde{i}_N \tilde{j}_N - \tilde{i}_{N-1} \tilde{j}_{N+1}) - n \tilde{i}_{N-1} \tilde{j}_{N+1} (\bar{v} z + \bar{z} v ) + \nonumber \\
& + n \left ( \tilde{i}_{N-1} \tilde{j}_{N+2} |v|^2 + \tilde{i}_{N-2} \tilde{j}_{N+1} |z|^2 \right ),
\end{align}
where we write $\tilde{i}_N = \tilde{i}_{\vec{n}}$, $\tilde{j}_N =
\tilde{j}_{\vec{n}}$. The building blocks are
\begin{align}
\tilde{i}_{\alpha} & = \frac{1}{\alpha!} \int_0^\infty d\rho e^{-\rho} (\rho+n|z|^2)^{\alpha}, \label{igindef} \\
\tilde{j}_{\beta} & = - \frac{(\beta-1)!}{2\pi i} \oint_{\Gamma} dp \frac{e^p\ln p}{(p + n|v|^2)^{\beta}}. \label{jgindef}
\end{align}
The bosonic block, when compared to Eq.~\eqref{ijreg}, lacks the term
$\gamma+\Gamma(0,p)$ since this contribution vanishes in the
generating function \eqref{rlrreg}, as can be seen by a symbolic
calculation. This observation holds more generally, not only in
this simplest case. Directly from the definitions, we derive the
iterative formulas
\begin{align*}
\tilde{i}_{\alpha} &= \tilde{i}_{\alpha-1} +
(n|z|^2)^{\alpha}(\alpha!)^{-1}, \\
\tilde{j}_{\beta} &=
\tilde{j}_{\beta+1} - (\beta-1)!  (n|v|^2)^{-\beta} e^{-n|v|^2}
\tilde{i}_{\beta-1}(v)
\end{align*}
and use them to re--express the generating function
\begin{align}
\label{ginratio}
\tilde{\mathcal{R}}_{G} = & n \tilde{i}_{N-1} \tilde{j}_{N+1} |v-z|^2 + \nonumber \\
& + \frac{e^{-n|v|^2}}{|v|^{2N}} \left (\tilde{i}_{N-1}(z)|v|^{2N} - \tilde{i}_{N-1}(v) |z|^{2N} \right ),
\end{align}
where we have written out explicitly the argument of $\tilde{i}$ to
avoid confusion.  At this point we observe that the generating
function vanishes for $z=v$, $\tilde{\mathcal{R}}_G=0$. It is thus
evident that the derivative formula \eqref{normalrho} only produces
contributions due to the second term. Lastly, by using
$\partial_{\bar{z}} \tilde{i}_\alpha = n z \tilde{i}_{\alpha-1}$ and
$\partial_v \tilde{j}_\beta = - n \bar{v} \tilde{j}_{\beta+1}$, we
recover the well--known formula 
\begin{align}
\label{ginspectr}
	\rho_G = \frac{n}{N\pi} e^{-n|z|^2} \sum_{k=0}^{N-1} \frac{(n|z|^2)^k}{k!},
\end{align}
for the spectral density, which often appears for $n=N$.

\subsection{Non--normal rank--1 $S$ and $L=R=1$}

A major reason to study models of the type~\eqref{meq} is the issue of
spectral stability. --- How far do the eigenvalues of $S+Y$ spread
around the eigenvalues of $S$ for a small perturbation $Y$. This is
especially interesting for finite rank sources $S$ where extremal (or
outlier) eigenvalues emerge from the eigenvalue sea of the matrix
$Y$. This phenomenon was studied in a Hermitian
\cite{Pec2005:HERMSMALLRANK,CDF2009:HERMRANKDEFORM,BN2011:HRANKPERTURB}
as well as a non--Hermitian
\cite{Tao2011:OUTLIERSNHSPECTRUM,TVK2010:CIRCULARLAW,OR2014:LOWRANKPERTURB}
setting. Here, we examine how the normal or non--normal character of
the source influences the eigenvalue distribution. We consider a
rank-one source of the form
 \begin{align}
	S = \alpha \ket{n} \bra{m},
\end{align}
for complex parameter $\alpha$ and bras (kets) $\bra{m}$ ($\ket{n}$) denoting the canonical matrix basis -- the source matrx $S$ has one non--zero element $\alpha$ placed on the off--diagonal. For the sake of simplicity we choose the trivial
variance structure $L=R=1$. After a fair amount of algebra we find the
result
\begin{align}
\label{rnn}
	\mathcal{R}_{NN} =  R_0 + |\alpha|^2 R_1 + |\alpha|^4 R_2 + |\alpha|^6 R_3 + |\alpha|^8 R_4
\end{align}
for the generating function. The formulas for the $R_i$'s are lengthy and thus were explicitly given only in the App.~B. Although the terms in Eq.~\eqref{rnn}
turn out to lack structure, they are still assembled
from the bosonic and fermionic building blocks similar to Eq.~\eqref{idef},
\begin{align}
  & i_{k,l}(z,w) = \frac{(-1)^k}{n^{k+2l+1}} e^{-n|w|^2} \int_0^\infty d\rho e^{-\rho} I_0(2\sqrt{n\rho}|w|) \times \nonumber \\
  & \times(\rho + n|z|^2)^k (\rho+nk^+_z)^l(\rho+nk^-_z)^l, 
\label{nnidef}
\end{align}
and Eq.~\eqref{jdef0},
\begin{align}
& j_{q,r}(v,u) = \frac{2}{i\pi} e^{n|u|^2}\int_{-\infty}^\infty dg g_- e^{-ng_-^2} K_0(2in|u|g_-) \times \nonumber \\
 & \times (g_-^2-|v|^2)^{-q}(g_-^2-k_v^+)^{-r}(g_-^2-k_v^-)^{-r}, 
\label{nnjdef}
\end{align}
where $k_x^\pm = \frac{1}{2} \left ( |\alpha|^2 + 2 |x|^2 \pm
  |\alpha|\sqrt{4|x|^2+|\alpha|^2} \right )$. By investigating the terms in each of the $R_i$'s, we
find the conditions $l=-1,0,1$, $k\geq 0$ and $q+r \geq 1$, $r=1,2,3$.
for the indices of $i_{k,l}$ and $j_{q,r}$, respectively.  We employ
the same regularization steps as in Sec.~\ref{secnormal}, obtain the
generating function $\tilde{\mathcal{R}}_{NN}$ and construct the
regularized fermionic block
\begin{align}
  & \tilde{i}_{k,0} = \frac{(-1)^k k!}{n^{k+1}}(\tilde{i}_{G})_k, \label{l0}\\
  & \tilde{i}_{k,1} = \tilde{i}_{k+2,0} - |\alpha|^2(\tilde{i}_{k+1,0}+|z|^2\tilde{i}_{k,0}), \label{l1}\\
  & \tilde{i}_{k,-1} = \frac{(-1)^k k!}{(k^+_z-k^-_z) n^k} \times \nonumber \\
  & \times \sum_{l=0}^k \frac{(n|z|^2)^l}{l!}\Big [
  U_{1,1+l-k}(nk^-_z) - U_{1,1+l-k}(nk^+_z) \Big ] ,
\label{lm1}
\end{align}
where $\tilde{i}_{G}$ is the Ginibre block of Eq.~\eqref{igindef} and
$k\geq 0$. We relegate the derivation of Eq.~\eqref{lm1} to the
App.~B. The bosonic block reads
\begin{align}
\tilde{j}_{q,r} = - \frac{(-n)^{q+2r-1}}{2\pi i} \oint_\Gamma \frac{dp e^p\ln p}{(p+n|v|^2)^q(p+nk^-_v)^r(p+nk^+_v)^r},
\end{align}
where $q \geq 0, r \geq 1$ and the contour $\Gamma$ encircles both
$-n|v|^2$ and $-nk^\pm_v$. Lastly we obtain the formulas for $q=-1,-2$,
\begin{align}
	\tilde{j}_{-1,2} = & \frac{1}{2} \left ( \tilde{j}_{0,2_-} + \tilde{j}_{0,2_+} + |\alpha|^2 \tilde{j}_{0,2} \right ), \\
	\tilde{j}_{-1,3} = & \frac{1}{2} \left ( \tilde{j}_{0,3_-} + \tilde{j}_{0,3_+} + |\alpha|^2 \tilde{j}_{0,3} \right ), \\
	\tilde{j}_{-2,3} = & \frac{1}{4} \left ( \tilde{j}_{0,3_{--}} + 2 \tilde{j}_{0,3_{+-}} + \tilde{j}_{0,3_{++}} + |\alpha|^4 \tilde{j}_{0,3} \right . + \nonumber \\
	& \left . + 2|\alpha|^2 (\tilde{j}_{0,3_+}+\tilde{j}_{0,3_-}) \right ), 
\end{align}
where the subscripts $\pm$ indicate that the underlying multiplicity
vector $\vec{x} = (q,r-1,r)$ is applied with decrement to the source
at $nk^\pm_v$.

Finally, we obtain the spectral density \eqref{rhodef} analytically
and plot it in Fig. 3. To facilitate a comparison, we juxtapose it
with the analogous results for the case of a rank--one normal source
$S$ and for the Ginibre case \eqref{ginspectr}. A non--normal source
$S$ (third row in Fig. 3) does not produce, on average, outlier
eigenvalues in the spectrum, in contrast normal source $S$ (second row
in Fig. 3) where we find an island around $\alpha=10$. Instead, in the
non--normal case we observe something like a blow--up of the spectral
bulk. The first row in Fig. 3 is devoted to the case of a vanishing
source, $S=0$. Near $z=0$ both, normal and vanishing source,
produce similarly shaped spectral densities --- the only difference
between these cases is the presence or absence of the finite--rank island.
\begin{figure}[ht!]
	\centering
	\includegraphics[width=.5\textwidth]{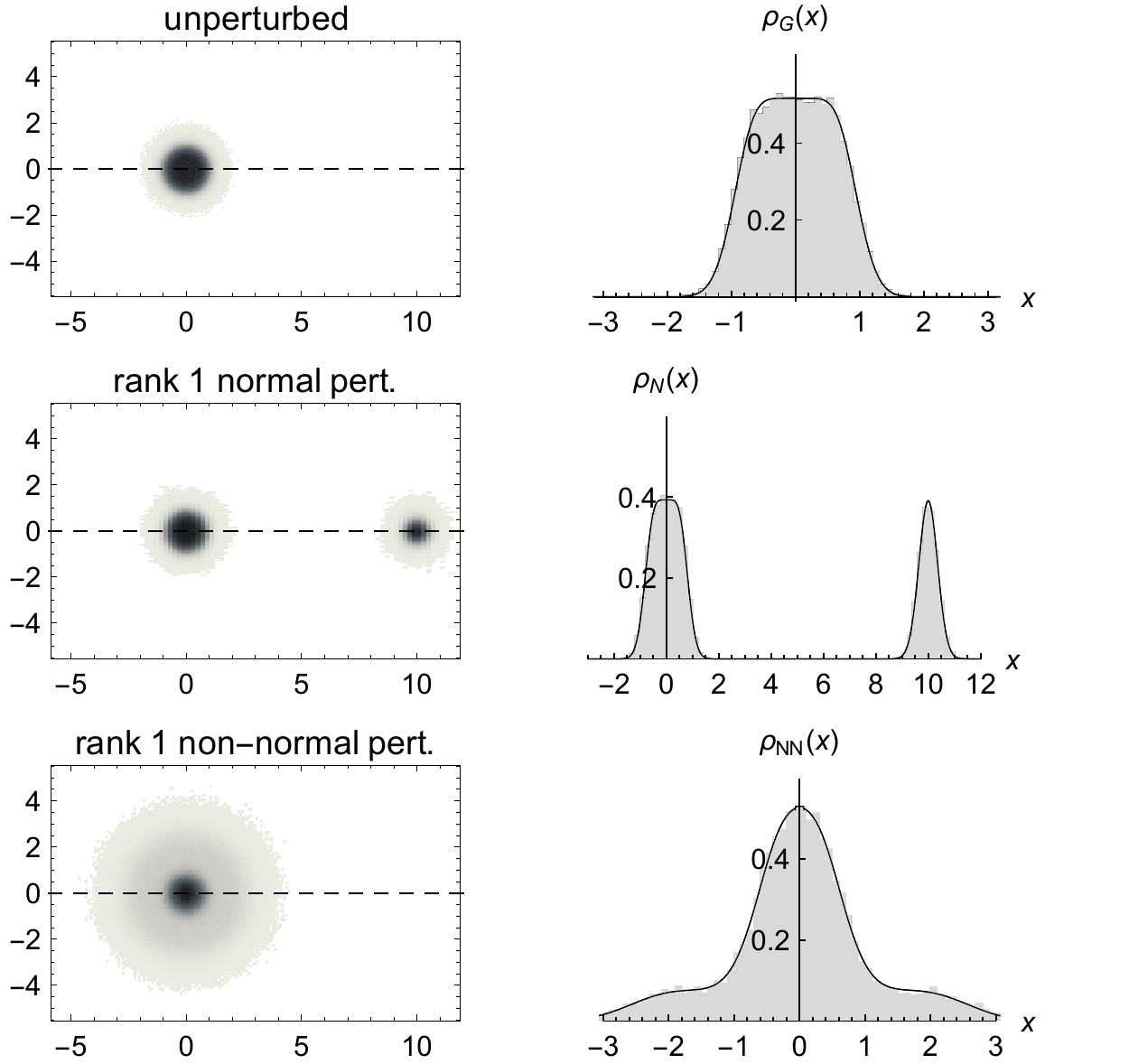}
        \caption{Left hand side: complex plane of eigenvalues, from
          top to bottom for: unperturbed $S=0$ (Ginibre), normal
          perturbation $S=10 \ket{1} \bra{1}$ and non--normal
          perturbation $S=10\ket{2}\bra{1}$. Right hand side:
          numerical simulations and analytical results for the
          spectral densities $\rho_G,\rho_N$ and $\rho_{NN}$ along the
          real axis line (dashed lines on the left hand side). 
          Numerical simulations are for matrices of size $N=4$,
          $\alpha=10$, we set $n=N$.}
\end{figure}

\subsection{Spectrum of $M^{-1}$}

As a last application we discuss how to infer somewhat gratuitously
the spectrum of $(S+X)^{-1}$ from the results for the spectrum of
$S+X$. For simplicity we deal with a normal source $S$ only and set
$L=R=1$. To this end we define a generating function
$\mathcal{R}_{-1}$ for the inverse as
\begin{align}
	\mathcal{R}_{-1}(Z,V) & = \left < \frac{\det (Z-\mathcal{M}_{-1})}{\det (V-\mathcal{M}_{-1})} \right > = \nonumber \\
	& = \frac{\det Z}{\det V} \mathcal{R}_{1,1}\left (Z',V' \right ) ,
\end{align}
and relate it to the generating function \eqref{ratiodef} previously
considered. The matrices $\mathcal{M}_{-1} = \left ( \begin{matrix} 0
    & M^{-1} \\ M^{\dagger,-1} & 0 \end{matrix} \right )$ and $Z',V'$
are rearranged versions of the inverse matrices $Z^{-1},V^{-1}$ of
Eq.~\eqref{zvdef},
\begin{align}
  X' = \left ( \begin{matrix} (X^{-1})_{22} & (X^{-1})_{21} \\
      (X^{-1})_{12} & (X^{-1})_{11} \end{matrix} \right ), \quad X =
  Z,V .
\end{align} 
We thus conclude that the whole calculation discussed in
Sec.~\ref{secnormal} can be repeated with only making the
replacements $w \to -w G_{zw}, z \to \bar{z} G_{zw}, u\to
-u G_{vu}$ and $v \to \bar{v} G_{vu}$ with $G_{xy} = (|x|^2 +
|y|^2)^{-1}$.
We again conduct the regularization procedure and eventually find that
only the source matrix of Eq.~\eqref{alphadef} is modified according
to
\begin{align*}
\alpha_{xy} \to (\alpha^{-1})_{xy} = \alpha_{x^{-1}y^{-1}} = (\bar{x}^{-1}\textbf{1}_N - S^\dagger)(y^{-1}\textbf{1}_N - S),
\end{align*}
The regularized ratio for the problem of finding the spectrum of
$(S+X)^{-1}$ reads
\begin{align}
\label{ratioinverse}
	& \tilde{R}_{-1} = \left ( \frac{|z|^2}{|v|^2} \right )^{|\vec{n}|} \tilde{\mathcal{R}}_{1,1} \left [ \alpha_{xy} \to (\alpha^{-1})_{xy} \right ] ,
\end{align}
where the generating function $\tilde{\mathcal{R}}_{1,1}$ is that of
Eq.~\eqref{rlrreg} and the constituent fermionic and bosonic blocks
\eqref{ijreg} are affected accordingly. In particular, we calculate the
spectral density for an inverse matrix $X^{-1}$ as
\begin{align}
	\rho_{G,-1} = \frac{n e^{-\frac{n}{|z|^2}}}{N\pi |z|^{4}}\sum_{k=0}^{N-1} \frac{1}{(k)!}\left ( \frac{n}{|z|^2} \right )^k,
\end{align}
obtained from Eq.~\eqref{ginratio}. This formula was also found in a
recent work on the product of matrices \cite{ARRS2016:PRODINV}.
\begin{figure}[ht!]
	\centering
	
	\includegraphics[width=.5\textwidth]{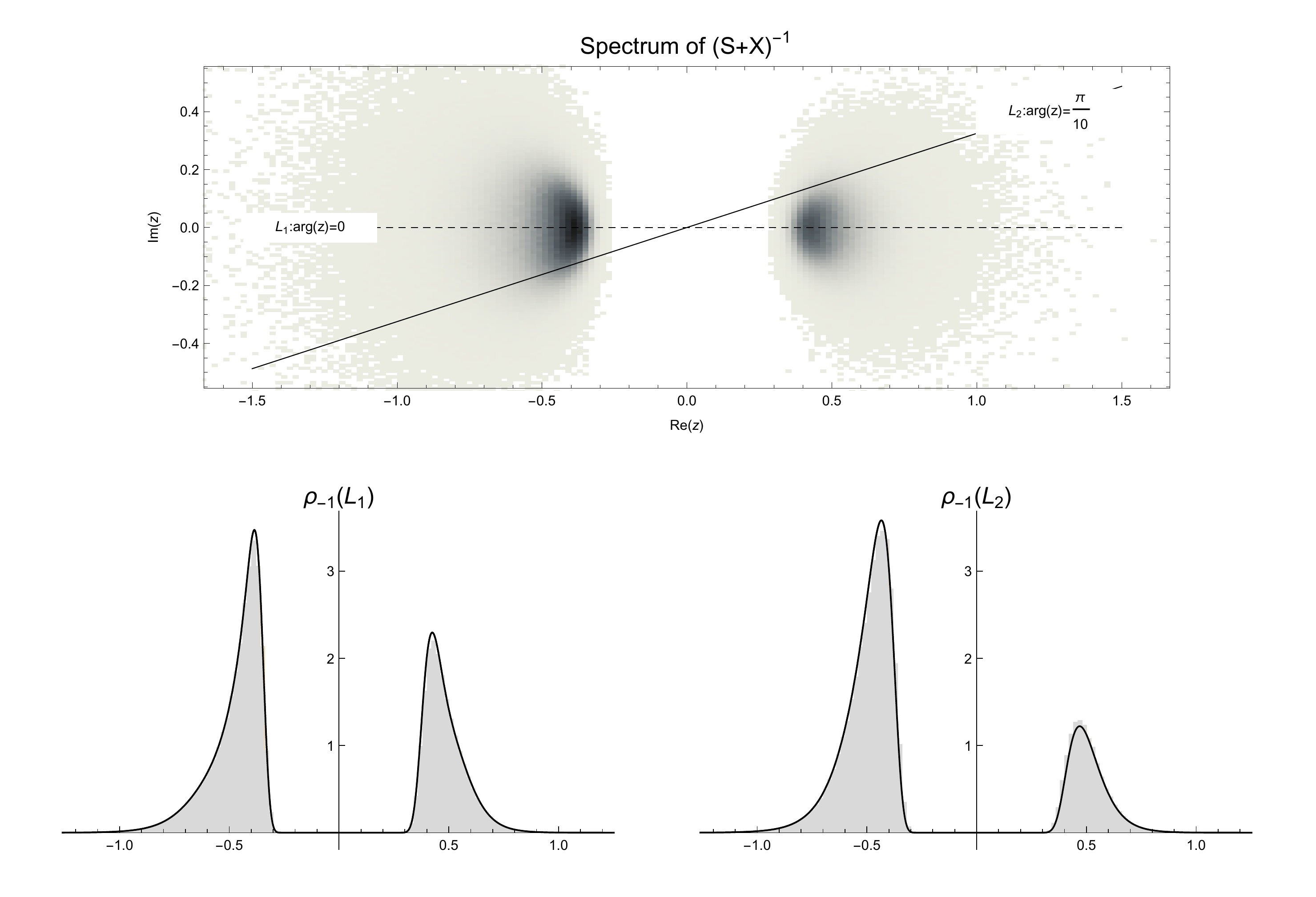}
      \caption{A numerical simulation along with analytic spectral density plots of matrix $(S+X)^{-1}$ along two straight lines $L_1$ and $L_2$ for an external source setup as $S=(-2,2)$ with multiplicities $\vec{n} = (4,2)$.}
\end{figure}
In Fig. 4, the spectral density of $(S+X)^{-1}$ is depicted as
calculated from the generating function \eqref{ratioinverse} for
non--zero external source $S$.
 
\section{Conclusions}
\label{sec4}

We have calculated exact spectral densities for a class of complex
random matrix models of the form $M=S+LXR$ consisting of a noise part
$X$ and structure parts $S,L,R$. We found two--fold integral formulas
for arbitrary structural matrices. In greater detail, we investigated
the case of a normal source matrix $S$ and arbitrary diagonal matrices
$L,R$ which are of particular interest. The resulting formulas 
are of a remarkably succinct form. We confirmed our analytical results 
by numerical simulations.

We showed how the presence or absence of the normality condition for
$S$ leads to a qualitatively different behaviour of the eigenvalue
densities. Our study was focused mainly on the finite rank source
matrices where analytical solutions proved tractable. For a
non--normal source, the most interesting feature is the lack of
outliers, \textit{i.e.}, extreme values in the averaged spectral
density. However, when imposing the normality condition on the source
matrix $S$, the outliers are clearly present in the spectral density.

Lastly, we looked at the problem of finding spectra of an inverse
matrix $M^{-1}$ which, by using the approach in this paper, proved to
be trivially connected to the spectrum of $M$.

Among the open problems in the context of our study, the question
remains on whether the normal vs.~non--normal dichotomy has any
counterpart relevant for applications. Secondly, the information on
eigenvectors is encoded in the objects of study but was, due to the
approach taken, completely omitted in our present work. Thirdly,
issues related to universality seem feasible within our approach 
and are certainly worth future investigation.

\section{Acknowledgements}

We gratefully acknowledge fruitful conversations with M.~Kieburg,
M.~Nowak and T.~Wirtz. JG appreciates the hospitality of
Duisburg--Essen University where part of this work was done in the
framework of the Erasmus+ exchange programme. JG also acknowledges the
support of the Grant DEC-2011/02/A/ST1/00119 of the Polish National
Centre of Science.

\appendix
\section{Derivation of \eqref{jdef}}
\label{appj}
We start from equation \eqref{jdef0}:
\begin{align}
j_{\vec{m}}(v,u) & = \frac{2in}{\pi}\prod_{i=1}^{k} \frac{(m_i-1)!}{(-n)^{m_i}} e^{n|u|^2} J_{\vec{m}}(v,u), \label{jintro} \\
J_{\vec{m}}(v,u) & = \int_{-\infty}^\infty dg g_- e^{-ng_-^2} K_0(2in|u|g_-) \prod_{i=1}^{k}(g_-^2 - \alpha^i_{vv})^{-m_i}, 
\end{align}
where we set $d(\vec{m})=k$ for brevity. By Lagrange interpolation formula we find:
\begin{align*}
	& \prod_{i=1}^k \left (g_-^2 - \alpha^i_{vv} \right )^{-m_i} =  \lim_{\gamma_1...\gamma_k \to 1} \sum_{l=1}^{k} \mathcal{D}_l (g_-^2 - \gamma_l \alpha^l_{vv})^{-1},
\end{align*}
with the operator $\mathcal{D}_l$ defined as
\begin{align*}
	& \mathcal{D}_l = \prod_{i=1}^k \frac{(\alpha^i_{vv})^{1-m_i}}{(m_i-1)!} \frac{d^{m_i-1}}{d\gamma_i^{m_i-1}}  \prod_{j=1(\neq l)}^k (\gamma_l \alpha^l_{vv} - \gamma_j \alpha^j_{vv})^{-1},
\end{align*}
So that the whole integral $J_{\vec{m}}$ is expressed as
\begin{align}
\label{joc}
	J_{\vec{m}} = \lim_{\gamma_1...\gamma_k \to 1} \sum_{l=1}^k \mathcal{D}_l C_l.
\end{align}
From now on we focus on the integral $C_l$:
\begin{align}
\label{intK0}
	C_l = \int_{-\infty}^\infty dg \frac{g_- e^{-n g_-^2}}{g_-^2 - \alpha^l_{vv} \gamma_l} K_0 (2 n i |u| g_-),
\end{align}
We re-introduce the representation $K_0(2ni|u|g_-) = \int_0^\infty ds \exp \left ( -2ni|u| g_- \cosh s \right )$ and compute:
\begin{align}
	C_l & = \frac{1}{2\sqrt{\gamma_l \alpha^l_{vv}}} \int_0^\infty ds \left (I_+(s) - I_-(s) \right ) ,\\
	I_\pm(s) & = \int_{-\infty}^\infty dg \frac{f(g_-,s)}{g - (\pm\sqrt{\gamma_l \alpha^l_{vv}}+i\epsilon) },
\end{align}
with $f(x,s) = x e^{-nx^2-2ni|u|x\cosh s}$. The integrals $I_\pm$ are calculable by Sokhotski--Plemelj formula:
\begin{align}
\label{id0}
	I_\pm(s) = i \pi f(\pm \sqrt{\gamma_l \alpha^l_{vv}},s) + \textrm{PV} \int_{-\infty}^\infty \frac{dx f(x,s)}{x-(\pm \sqrt{\gamma_l \alpha^l_{vv}})} .
\end{align}
The second part is the Hilbert transform \cite{Kin2009:HILBTRANS}:
\begin{align}
	& \frac{1}{\pi} \textrm{PV} \int_{-\infty}^\infty dy \frac{y e^{-ay^2-by}}{y-x} = \frac{1}{\sqrt{a\pi}} e^{b^2/4a} + \nonumber\\
	& + i x e^{-x^2a - xb}  \text{erf} \left (\frac{i}{2\sqrt{a}}(b+2ax) \right ) . \label{id1}
\end{align}
Lastly, we need the identity:
\begin{align}
	& \int_x^\infty dt e^{-a^2 t^2 - b^2/t^2} = \nonumber \\
	& = \frac{\sqrt{\pi}}{4a} \left ( e^{2ab} \text{erfc}(ax+b/x) + e^{-2ab} \text{erfc}(ax-b/x) \right ), \label{id2}
\end{align}
valid for $x>0$. Combining the formulas of \eqref{id0}-\eqref{id2} result in
\begin{align}
C_l = & 2 i \sqrt{\pi n} |u| e^{-n\alpha^l_{vv} \gamma_l} \times \nonumber \\
& \times \int_0^\infty ds \int_1^\infty dt \cosh s~ e^{\frac{n \alpha^l_{vv} \gamma_l}{t^2}- n|u|^2 t^2 \cosh^2 s},
\end{align}
In the next step we integrate over $s$ and change variables $t^2 = \tau+1$:
\begin{align}
C_l = \frac{i \pi}{2} \int_0^\infty d\tau \frac{1}{\tau+1} e^{-n|u|^2(\tau+1) - n\gamma_l \alpha^l_{vv} \frac{\tau}{\tau+1}}.
\end{align}
We introduce a succinct contour integral representation:
\begin{align*}
	\lim_{\gamma_1...\gamma_k \to 1}\sum_{l=1}^k \mathcal{D}_l e^{-n\gamma_l \alpha^l_{vv} \frac{\tau}{\tau+1}} = \frac{1}{2\pi i} \oint_{\Gamma_s'} dq \frac{e^{-nq\frac{\tau}{\tau+1}}}{\prod_{i=1}^k(q-\alpha^i_{vv})^{m_i}} ,
\end{align*}
where the contour $\Gamma_s'$ encircles all $\alpha^i_{vv}$'s counter-clockwise. This formula is a part of \eqref{} which, after changing $p=-nq$ is equal to:
\begin{align}
	J_{\vec{m}} = & \frac{i\pi}{2} (-n)^{|\vec{m}|-1}e^{-n|u|^2} \times \nonumber \\
	& \times \frac{1}{2\pi i} \int_0^\infty d\tau \oint_{\Gamma_s} dp \frac{1}{\tau+1} \frac{e^{-n|u|^2 \tau + \frac{p\tau}{\tau+1}}}{\prod_{i=1}^k (p+n\alpha^i_{vv})^{m_i}}, \label{jfin}
\end{align}
with appropriately modified contour $\Gamma_s$.
Lastly, we use an integral representation of the Tricomi confluent hypergeometric function:
\begin{align*}
	\int_0^\infty d\tau \frac{1}{\tau+1} e^{-n|u|^2 \tau + \frac{p\tau}{\tau+1}} = \sum_{k=0}^\infty U_{k+1,1}(n|u|^2) p^k ,
\end{align*}
and combine it with \eqref{jintro} and \eqref{jfin}:
\begin{align*}
j_{\vec{m}} = \frac{\prod_{i=1}^{|\vec{m}|} (m_i-1)!}{2\pi i}\oint_{\Gamma_s} dp \sum_{k=0}^\infty \frac{U_{k+1,1}(n|u|^2)p^k}{\prod_{i=1}^{|\vec{m}|} \left (p+n \alpha^i_{vv} \right )^{m_i}} ,
\end{align*}
which is exactly the formula \eqref{jdef}.
\section{Details of non-normal $S$ case}
\label{appnn}
The ratio for non-normal case is given by \eqref{rnn} with $R_i$ terms:
\begin{widetext}
\begin{align*}
R_0 & = 2(V i_{N-3,1} j_{N,1} + Z i_{N,-1} j_{N-3,2}) + 6 (V i_{N-1,0} j_{N-4,3} + Z i_{N-4,1} j_{N-1,1}) - 4V j_{N-2,2} \delta_1^+ - 4Z i_{N-2,0} \sigma_1^+ + \\
& + N^2 \left [ j_{N-1,1} \Delta_{N-3,1}^{Z} + V i_{N-3,1} j_{N,1} \right ] + n d_1 \left [ (N-2) j_{N-1,1} i_{N-3,1} + 2i_{N-1,0} j_{N-3,2} \right ] - n^2 i_{N-2,1} j_{N-2,1} + \\
& + N \left [ 2Vj_{N-2,2} \delta_1^+ - 2Z j_{N-1,1} \delta_2 + 2j_{N-3,2} \Delta_{N-1,0}^{Z} - 2 j_{N-1,1} \Delta_{N-3,1}^{Z} - Z i_{N-4,1} j_{N-1,1} - 3V i_{N-3,1} j_{N,1} \right ], \\
R_1 & = -N \left [ \delta_1^- \Sigma_{N-2,2}^{V} + \Delta_{N-2,0}^{Z} \sigma_1^- \right ] + n \left [ 2 \Delta_{N-1,0}^Z \Sigma_{N-3,2}^{V} + d_2 i_{N-2,0} j_{N-2,2} \right ] + i_{N-1,0} (2V j_{N-1,2} + 3 j_{N-4,3}) + \\
& + d_1 \left [ 2N j_{N-2,2} \Delta_{N-2,0}^{Z} + i_{N-2,0} (4Vj_{N-3,3} - N j_{N-2,2}) + i_{N-2,0} j_{N-4,3} - i_{N,-1} j_{N-2,2}+ V(N-2) i_{N-2,0} j_{N-1,2} + \right .\\
& - Z (N+2) i_{N-3,0} j_{N-2,2} \Big ] + 2V j_{N-4,3} \delta_3 - Z i_{N,-1} \sigma_2 - 2 i_{N-3,1} \Sigma_{N-2,2}^{V} - 2 j_{N-1,1} \Delta_{N-2,0}^{Z} - 2Z j_{N-3,2} \Delta_{N-1,-1}^{Z} +\\
&  + 2V i_{N-1,0} \Sigma_{N-3,3}^{V} + j_{N-3,2}(2Z i_{N-3,0} + i_{N,-1}) - (Vi_{N-2,0} - Z i_{N,-1}) j_{N-4,3}, \\
R_2 & = d_1 \left [ \Delta_{N-1,-1}^{Z} j_{N-2,2} + (i_{N-2,0} - 2 i_{N,-1}) \Sigma_{N-3,3}^{V} \right ] + 2(N-2) \Delta_{N-2,0}^{Z} \Sigma_{N-2,2}^{V} + V i_{N-2,0} \Sigma_{N-3,3}^{-V} + \\
& - 2 (Z+V) j_{N-4,3} \Delta_{N-1,-1}^{Z} - \Sigma_{N-3,2}^-(Z) \Delta_{N-1,-1}^{Z} - i_{N-1,0} \Sigma_{N-3,3}^{V} + j_{N-2,2} i_{N-2,0} - \delta_3 \sigma_2, \\
R_3 & = -\delta_3 \Sigma_{N-3,3}^{V} + \Delta_{N-1,-1}^Z \left [ \sigma_2 + 2 d_1 \Sigma_{N-3,3}^V \right ], \\
R_4 & = \Delta_{N-1,-1}^Z \Sigma_{N-3,3}^V,
\end{align*}
\end{widetext}
where $V=|v|^2, Z=|z|^2, d_1 = \bar{z} v + z \bar{v}, d_2 = (\bar{z} v)^2 + (z \bar{v})^2$ and the notation reads
\begin{align*}
& \delta_1^\pm = i_{N-1,0} \pm i_{N-3,1}, \quad \delta_2 = i_{N-4,1} - i_{N-2,0}, \\
& \delta_3 = i_{N,-1} - i_{N-2,0}, \\
& \sigma_1^\pm = j_{N-3,2} \pm j_{N-1,1}, \quad \sigma_2 = j_{N-2,2} - j_{N-4,3}, \\
& \Delta_{x,y}^z = i_{x,y} + z i_{x-1,y}, \quad \Sigma_{x,y}^z = j_{x,y} + z j_{x+1,y}.
\end{align*}
Now we turn to the calculation of regularized bosonic block $\tilde{i}_{k,-1}$ of \eqref{lm1}. We start from the definition \eqref{nnidef}:
\begin{align*}
	& i_{k,-1} = \frac{(-1)^k}{n^{k-1}} e^{-n|w|^2} \times \nonumber \\
	& \times \int_0^\infty d\rho e^{-\rho} I_0(2\sqrt{n\rho}|w|) \frac{ (\rho + n|z|^2)^k}{(\rho + nk^+_z)(\rho + nk^-_z)}.
\end{align*}
Firstly, we express the denominator as an integral:
\begin{align*}
	\frac{1}{(\rho + nk^+_z)(\rho + nk^-_z)} = \frac{1}{2n\delta k} \int_0^\infty dp e^{-p\rho-pn k_0} \sinh (pn\delta k),
\end{align*}
\\
with $k^\pm_z = k_0 \pm \delta k$. We consider the integral:
\begin{align*}
	& \mathcal{I}(p) = \int_0^\infty d\rho e^{-(1+p)\rho} (\rho + n|z|^2)^k I_0(2\sqrt{n\rho}|w|) = \\
	& = e^{\frac{n|w|^2}{p+1}}\frac{(n|z|^2)^k k!}{p+1}\sum_{l=0}^k \frac{(n|z|^2(p+1))^{-l}}{(k-l)!} L_l\left ( - \frac{n|w|^2}{p+1} \right ),
\end{align*}
and obtain the formula for $i_{k,-1}$:
\begin{align*}
i_{k,-1} = \frac{(-1)^k}{2n^{k} \delta k} e^{-n|w|^2} \int_0^\infty dp e^{-pnk_0} \sinh (pn\delta k) \mathcal{I}(p).
\end{align*}
\\
It gets simplified in the regularization $w\to 0$ limit:
\begin{align*}
	& \tilde{i}_{k,-1} = \frac{(-1)^k k!}{2n^k\delta k} \times \\
	& \times  \sum_{l=0}^k \frac{(n|z|^2)^l}{l!}\Big [ U_{1,1+l-k}(nk^-_z) - U_{1,1+l-k}(nk^+_z) \Big ],
\end{align*}
thus reproducing the equation \eqref{lm1}.
\bibliography{krkbib2015}{}
\bibliographystyle{plain}

\end{document}